\documentclass[12pt]{article}
\usepackage{amsmath}
\usepackage{graphicx}
\usepackage{enumerate}
\usepackage{natbib}
\usepackage{url} % not crucial - just used below for the URL 

\usepackage{amsthm}

\usepackage[nottoc]{tocbibind}

\usepackage[colorlinks =true,linkcolor=blue,urlcolor=blue,citecolor=blue]{hyperref}

\usepackage{ragged2e}

\usepackage{subcaption}
\usepackage{url}
\usepackage[utf8]{inputenc}
\usepackage{tabulary}
\usepackage{setspace}
\usepackage{authblk}
\usepackage{natbib}
\usepackage{hyperref}
\usepackage{float}
\usepackage{multirow}
\usepackage{graphicx}
\usepackage{adjustbox}
\usepackage{caption}
\usepackage{booktabs}
\usepackage{amsfonts}
\usepackage{amssymb}
\usepackage{amsxtra}
\usepackage{amsthm}
\usepackage{mathrsfs}
\usepackage{bbm}
\usepackage{enumerate}
\usepackage{enumitem}
\usepackage{xcolor}
\usepackage{lscape}
\usepackage{pgfplots}

\usepackage{enumitem}

\newtheorem{assumption}{Assumption}
\theoremstyle{definition}

\newtheorem{prop}{Proposition}

\newtheorem{corollary}{Corollary}
\newtheorem{remark}{Remark}

\newcommand{\bitemsize}{\begin{itemize}}
\newcommand{\eitemsize}{\end{itemize}}
\newcommand{\be}{\begin{equation*}\begin{aligned}}  
\newcommand{\ee}{\end{aligned}\end{equation*}}
\newcommand{\benum}{\begin{enumerate}}
\newcommand{\eenum}{\end{enumerate}}
\newcommand{\bmat}{\begin{bmatrix}}
\newcommand{\emat}{\end{bmatrix}}

\newcommand{\pr}{\mathrm{P}}

\newcommand{\var}{\mathrm{Var}}
\newcommand{\Var}{\mathrm{Var}}
\newcommand{\cov}{\mathrm{Cov}}

\newcommand{\E}{\mathrm{E}}
\renewcommand{\vec}[1]{\boldsymbol{#1}}

\newcommand\independent{\protect\mathpalette{\protect\independenT}{\perp}}
\def\independenT#1#2{\mathrel{\rlap{$#1#2$}\mkern2mu{#1#2}}}

\newcommand{\blind}{1}

\begin{document}

\def\spacingset#1{\renewcommand{\baselinestretch}%
{#1}\small\normalsize} \spacingset{1}

%%%%%%%%%%%%%%%%%%%%%%%%%%%%%%%%%%%%%%%%%%%%%%%%%%%%%%%%%%%%%%%%%%%%%%%%%%%%%%

\if1\blind
{
  \title{\bf Linear Regression in a Nonlinear World}
  \author{Nadav Kunievsky\thanks{Knowledge Lab, University of Chicago. This manuscript has benefited from numerous conversations with Natalie Goldshtein. It has also benefited from useful comments provided by \href{https://www.refine.ink/}{refine.ink}}.}
  \maketitle
} \fi

\if0\blind
{
  \bigskip
  \bigskip
  \bigskip
  \begin{center}
    {\LARGE\bf Linear Regression in a Nonlinear World}
\end{center}
  \medskip
} \fi

\bigskip
\begin{abstract}
The interpretation of coefficients from multivariate linear regression relies on the assumption that the conditional expectation function is linear in the variables. However, in many cases the underlying data generating process is nonlinear. This paper examines how to interpret regression coefficients under nonlinearity. We show that if the relationships between the variable of interest and other covariates are linear, then the coefficient on the variable of interest represents a weighted average of the derivatives of the outcome conditional expectation function with respect to the variable of interest. If these relationships are nonlinear, the regression coefficient becomes biased relative to this weighted average. We show that this bias is interpretable, analogous to the biases from measurement error and omitted variable bias under the standard linear model. 
\end{abstract}

\noindent%
\vfill

\newpage
% \spacingset{1.9} % DON'T change the spacing!
\onehalfspacing
\section{Introduction}
Multivariate linear regression is a fundamental tool across most scientific disciplines. Its main usage is to explore the relationship between different variables, assessing the average change in an outcome variable in response to an increase in the variable of interest (e.g \cite{wooldridge2015introductory}, \cite{weisberg2005applied}, \cite{greene2003econometric}, \cite{Cunningham2021Mixtape}, \cite{montgomery2012introduction}). For example, in environmental science, researchers may want to know what an increase in air quality implies for public health outcomes. To do this, researchers usually regress various outcomes on air quality measures, controlling for other variables like socioeconomic status and urbanization. They then interpret the regression coefficient on air quality as the change in health outcomes in response to a unit increase in air quality, holding the control variables fixed. 

Beyond describing correlations, multivariate linear regression plays a central role in causal analysis, especially in observational studies. When researchers want to estimate a causal effect, they often operationalize the conditional independence assumptions (\cite{pearl2009causality}, \cite{Cunningham2021Mixtape}, \cite{angrist2009mostly}) necessary for identifying causal relationships by regressing an outcome variable on the variable of interest and a set of control variables. The regression coefficient is then interpreted as the causal effect of a change in the variable of interest on the outcome.

Whether multivariate linear regression is used to learn about causal effects or conditional associations between variables, the clarity of interpretation largely hinges on the linearity of the conditional expectation function. If this function is indeed linear, the regression coefficient on the variable of interest coincides with its constant marginal effect, holding the other variables fixed. When the conditional expectation is nonlinear, however, regression coefficients generally no longer represent such a simple marginal effect.

This paper extends the findings of \cite{Yitzhaki1996} to clarify what multivariate regression recovers in this nonlinear setting. We introduce the \emph{Naive Regression-Weighted Effect (NRWE)} estimand: a weighted average of the derivative of the conditional expectation of the outcome with respect to the variable of interest, where the weights resembles conditional Yitzhaki's weights. We show that when the relationship between the variable of interest and the covariates is linear, the population regression coefficient on the variable of interest equals to the NRWE. Therefore, in this case, the coefficient can be interpreted as a particular regression-weighted average of conditional derivatives, with weights that resemble those in \cite{Yitzhaki1996} but are now defined cell-by-cell in the covariates and then averaged across them. When the relationship between the variable of interest and control variables is nonlinear, the coefficient departs from the NRWE in a systematic way: it equals a shrunk version of this weighted effect plus an interpretable bias term, analogous to classical measurement error and omitted-variable bias under a linear data-generating process.

We also show that the NRWE is of independent interest as a target estimand. First, as the NRWE is equal to the regression coefficient on the variable of interest, when the the relations between the variable of interest and controls is linear, it extremely easy to estimate.  Second, we show that the NRWE weights are essentially the unique way to represent the regression coefficient as a weighted average of partial derivatives of the conditional expectation. Third, among a broad class of average-derivative functionals that use well-behaved weights, the NRWE attains the lowest semiparametric efficiency, bound under homoskedasticity. This motivates independent interest in the NRWE as a target parameter from an efficiency perspective. Taken together, these properties justify using the NRWE as a natural lens through which to interpret the regression coefficient. 

There are several interpretations of linear regression coefficients in the literature. While our main antecedent is Yitzhaki’s (1996) univariate average-derivative result for simple regression (which we extend in Section \ref{NRWEDiscussion}), \cite{Yitzhaki1996} also discusses a separate multivariate decomposition: each coefficient in a multivariate regression can be expressed as a weighted sum of coefficients from simple regressions of the outcome on each control variable individually. This decomposition is less intuitive and departs from the standard “holding other variables constant” interpretation. Closest to our results, \cite{angrist1999empirical} shows that, in the case of a multi-value discrete treatment variable in a multivariate regression with fully saturated controls, the coefficient on the treatment variable can be interpreted as a weighted average of treatment effects across different treated groups. We extend this result to the continuous case and consider both settings in which the model is sufficiently rich to be viewed as saturated and settings in which the model is more restrictive and what bias this induces.

Additionally, recent econometric studies have focused on interpreting regression coefficients as weighted averages of treatment effects. For instance, \cite{GoldsmithPinkham2022Contamination} demonstrates that, when regressing an outcome on multiple treatment indicators, the coefficients on these variables are generally contaminated by the effects of other treatments. Likewise, the literature on difference-in-differences and event study designs (\cite{callaway2021difference}; \cite{RothSantAnna2022}; \cite{SunAbraham2021}; \cite{deChaisemartin2022DID}) discusses how these coefficients can be viewed as different weighted sums of heterogeneous treatment effects. \cite{blandhol2022tsls} characterize when TSLS has a credible “causal averaging” interpretation and when it instead mixes complier and non-complier contrasts with possibly negative weights. 

Most relevant to our work, \cite{ishimaru2024empirical}  formalizes how linear estimators implicitly average heterogeneous marginal effects using non-uniform weights. \cite{ishimaru2024empirical} shows that, with a continuous treatment and covariates, the OLS admit weighted-average representations of the underlying marginal effect function. In particular, the OLS estimand can be written as an average of marginal effects with covariate weights proportional to the conditional variance of the treatment (and normalized to integrate to one). Our paper make the OLS weights explicit and shows when the OLS coefficient equals this particular weighted average exactly and how departures from the linear specification generate a transparent misspecification term. Our paper clarifies which parts of the covariate distribution OLS emphasizes—namely, cells where the treatment is more variable—and provides guidance for empirical design and interpretation. Moreover, our paper provides an efficiency-based rationale for why the OLS weighting scheme may be appealing in its own right.

In this paper, we focus on the common case in the social sciences, where the variable of interest is continuous and the model includes control variables. We investigate how imposing a linear structure in estimation interacts with the underlying conditional expectation function. We then explore the biases and errors in interpretation that can arise when linearity assumptions are inappropriate.

\section{The Univariate Case}
We begin by considering the univariate case. Suppose that the underlying data generating process (DGP) is represented by the function:
$$
Y = g(T,\epsilon),
$$
where $Y$ is the outcome of interest, $T$ is a continuous variable of interest, $\epsilon$ represents unobserved variables that affects the outcome and may be correlated with $T$. $g(.)$ is the function that describes the DGP. A researcher interested in the effect of $T$ might estimate its coefficient in the linear model:
\begin{equation}
Y = \alpha + \beta T + u .
\end{equation}
In the univariate case, Yitzhaki's theorem (\cite{Yitzhaki1996}) provides a method to associate the population regression coefficient with the underlying DGP. Specifically, under certain regularity conditions, Yitzhaki shows that 
\begin{equation}\label{yithzhakiResult}
\beta = \int_{-\infty}^{\infty} \frac{\partial_ E[Y|t]}{\partial t} w(t) dt, 
\end{equation}
where $w(t) = \frac{E[T-E[T]|T>t]\pr(T>t)}{\var(T)}$ and $\int_{-\infty}^{\infty} w(t) = 1 $. These weights are maximized at $E[T]$ and are increasing on the left of the maximum value and decreasing on the right, assigning zero weight to the values at the boundaries of the support.\footnote{The weights can also be thought of a density function, but notice that the density is different than the density of $T$}  
Moreover, if  $\epsilon$ is independent of $T$, then the regression coefficient provides us with a positively weighted average of the average marginal causal effects of $T$ on $Y$, as 
\be
\frac{\partial E[Y|T=t]}{\partial t} =
E\bigg[\frac{\partial g(t,\epsilon)}{\partial t }\bigg].
\ee
where the expectations is taken over $\epsilon$, which implies then that $\beta$ is 
\begin{equation*}
\beta = \int_{-\infty}^{\infty} E\bigg[\frac{\partial g(t,\epsilon)}{\partial t }\bigg] w(t) dt, 
\end{equation*}

\section{The Naive Regression weighted Effect and Multivariate Regression}\label{NRWEDiscussion}
In this section, we extend Yitzhaki's result to the more generalized case, where we allow for additional control variables. Assume the DGP is now represented by the following two equations: 
\begin{align}
Y & = g(T,\vec{X},\upsilon), \\  
T & = h(\vec{X},\varepsilon),
\end{align}
where $\upsilon$ and $\varepsilon$ are unobserved variables that influence the outcome and the variable of interest's value, respectively, and $h$ and $g$ are the underlying causal functions that govern the DGP. 

When researchers want to learn on how a change in variable of interest, $T$, affects the expected outcome variable $Y$, while conditioning on $\vec{X}$, they often resort to using linear regression. Specifically, they may estimate the following linear model:\footnote{Throughout the analysis, we assume that $\vec{X}$ contains a constant.}
\begin{equation}
Y =  T\beta +  \vec{X}\gamma  + \epsilon. 
\end{equation}
If researchers do not want to assume that the DGP is linear, they often interpret the coefficient on the variable of interest, $\beta$, as an average of the marginal effect of $T$, holding $\vec{X}$ fixed. Naturally, one might wonder how Yitzhaki's theorem applies to this multivariate context and how we should interpret $\beta$ in this case.

To answer this question, we first define the "Naive Regression-Weighted Effect" (NRWE) as: 
\begin{equation}\label{nrwe_def}
NRWE = \E_{\vec{X}}\bigg[ \int_{-\infty}^{\infty}
\frac{\partial E[Y|t,\vec{X}]}{\partial t} w(t,\vec{X})  dt \bigg],
\end{equation}
where $w(T,\vec{X}) = \frac{E[T-E[T]|T>t, \vec{X}]\pr(T>t|\vec{X})}{E_X[\var(T|\vec{X})]}$. This parameter intuitively extends Yitzhaki's interpretation of $\beta$ from the univariate case to the multivariate case. To see that, notice that for each $\vec{X}$-cell, the numerator of the weights assigns the same weight that Yitzhaki's weights would assign in a regression of the outcome variable on $T$ at the particular value of $\vec{X}$. The denominator of the weights is simply the average over the conditional variance of $T$, which ensures that that $E_X[\int w(t,X)dt] =1 $ weights sum to 1, in a manner similar to Yitzhaki's original weights. 

\begin{remark}
Notice that under a sufficient conditional independence assumption, the NRWE captures a weighted average of causal effects. Specifically, assume that \(\upsilon\) is independent of \(T\) given \(\vec{X}\), i.e., \(T \independent \upsilon | \vec{X}\). Under this assumption, the NRWE can be expressed as:
\[
\E\left[  \int_{-\infty}^{\infty}\frac{\partial \E[Y|t,\vec{X}]}{\partial t} w(t,\vec{X}) \, dt \right] = 
 \E\left[ \E\left[\int_{-\infty}^{\infty} \frac{\partial g(t,\vec{X},\upsilon)}{\partial t}w(t,\vec{X})\, dt \mid \vec{X}\right]    \right].
\]
In this expression, the NRWE provides a weighted average of the causal marginal changes in the treatment variable \(T\). The inner expectation represents the marginal effect of changes in \(T\) on the outcome, holding \(\vec{X}\) fixed while averaging over the distribution of \(\upsilon\). The outer expectation then averages these effects across the distribution of \(\vec{X}\). Additionally, the implied weights of the marginal causal effects, \(w(t,\vec{X}) \Pr(\vec{X})\), are positive and sum to one. In this way, under the conditional independence assumptions, the NRWE effectively summarizes how changes in \(T\) causally influence \(Y\) across the population.
\end{remark}

\subsection{The Naive Regression-Weighted Effect and Multivariate Regression}

Proposition \ref{yitzhakiGeneralizedThm}, detailed below, shows that the population regression coefficient, $\beta$ in a multivariate regression, is equivalent to the $NRWE$ when the relationship between the controls and the variable of interest is linear. In other cases, it often yields a biased estimate in relation to the Naive Regression-Weighted Effect.

\begin{prop}[Multivariate Yitzhaki's Theorem]\label{yitzhakiGeneralizedThm}
Denote by $\pi$ the coefficients of $\vec{X}$ in the population regression of $T$ on $\vec{X}$. Denote by $\mu(\vec{X}) = E[T|\vec{X}]$ and denote the misspecification error by $\Delta(\vec{X}) = \mu(\vec{X}) - \pi \vec{X} $. Assume the first and second moments and conditional moments exist and that the conditional expectations $E[Y|T,\vec{X}]$ is differentiable with respect to $T$, then the regression coefficient on the variable of interest, $\beta$, in the population regression, $Y =  T\beta + \vec{X}\gamma + \epsilon $, is given by: 
\be
\beta & = \underbrace{\frac{\cov(Y,(T-\mu(\vec{X})) }{\var(T-\mu(\vec{X})) + \var(\Delta(\vec{X})) }}_{\text{Weighted Effect of $T$}}
 + \underbrace{\frac{\cov(Y,\Delta(\vec{X}))}{\var(T-\mu(\vec{X})) + \var(\Delta(\vec{X}))} }_{\text{Misspecification Bias}}
\\
 & = \underbrace{E_{\vec{X}}\bigg[ \int_{-\infty}^{\infty} \frac{\partial E[Y|T=t,\vec{X}]}{\partial t } \omega_{OLS}(t,\vec{X}) dt \bigg]}_{\text{Weighted Effect of $T$}} + \underbrace{\frac{\cov(Y,\Delta(\vec{X}))}{\E_{\vec{X}}[\var(T|\vec{X})] + \var(\Delta(\vec{X})) } }_{\text{Misspecification Bias}},
\ee    
where 
\be
\omega_{OLS}(t,x) & = 
\frac{\E[T-\mu(\vec{X})|T> t , \vec{X}]\pr(T>t|\vec{X})}{\E_{\vec{X}}[\var(T|\vec{X})] + \var(\Delta(\vec{X})) } \geq 0.
\ee
\end{prop}

The proof, detailed in the appendix, applies the Frisch-Waugh-Lovell Theorem and integration by parts.\footnote{\cite{angrist1999empirical} demonstrated for the discrete case, with a fully saturated regression, a similar equivalence between a discrete equivalent of the NRWE and the regression coefficient.} An immediate takeaway from Proposition \ref{yitzhakiGeneralizedThm} is that, generally, \(\beta\) does not equal the NRWE. This difference arises from two factors. First, the weights \(\omega_{OLS}(t,x)\), though resembling those in the NRWE, do not integrate to 1, which induces a bias analogous to the classical measurement error attenuation bias (e.g., \cite{wooldridge2015introductory}). This bias causes the effect of \(T\) to be attenuated in \(\beta\) relative to the NRWE. If the misspecification error variance is non-zero, \(\var(\Delta(\vec{X})) > 0\), the effect of \(T\) in \(\beta\) will be smaller than in the NRWE. 

The second source of bias in $\beta$ compared to the NRWE is the misspecification bias, driven by the covariance between the misspecification error and the outcome variable. To better understand this bias, we consider different DGPs. First, let us assume $g(T,\vec{X},\upsilon) = \beta T + \gamma \vec{X} + \upsilon$, and allow $h$ to be unrestricted. In this case, the outcome equation is correctly specified. By using the standard argument from the consistency of the OLS, we understand that the population regression coefficient equals the structural $\beta$, and, is therefore, trivially equal to the $NRWE$ parameter.\footnote{To see this through the lens of Proposition \ref{yitzhakiGeneralizedThm}, notice that:   
\be
\cov(Y,T-\mu(\vec{X})) + \cov(Y,\mu(X) - \pi \vec{X}) = \cov(Y,T - \pi \vec{X}) = \beta\cov(T,T-\pi X ).
\ee
Divide by the denominator to get $\beta$.
}
Next, let us consider the case in which $h$, the function governing the intensity of the variable of interest, is linear in $\vec{X}$. In this case, $\Delta(\vec{X})=0$ for all $\vec{X}$, and Proposition \ref{yitzhakiGeneralizedThm} shows that both the bias term and $\var(\Delta(\vec{X})) $ equal zero, which implies that $\beta$ equals the Naive Regression Weighted Effect. 

Finally, let us consider the case where both $g(T,\vec{X},\upsilon)$ and $h(\vec{X},\epsilon)$ are non-linear in $\vec{X}$. In this scenario, the population regression coefficient doesn't yield a weighted average of treatment effects. Instead, it gives a weighted average of the marginal effect of $T$ and an additional bias term, which is introduced by the correlation between the outcome variable $Y$ and $\Delta(X)$. These $\Delta(X)$ terms represent deviations of the conditional expectations from their best linear approximation.\footnote{Recall that the coefficients provide the best linear approximation to the conditional expectations. See, for example,(\cite{angrist2009mostly})} If these deviations are systematically correlated with the outcome variable, the estimate will be biased. If, however, we find that $\cov(Y,\Delta(\vec{X}))=0$ but $\var(\Delta(\vec{X})) \neq 0$, then the bias term in Proposition \ref{yitzhakiGeneralizedThm} becomes zero, but the measured effect is attenuated due to the variance of the measurement error in the denominator. Importantly, even if the control variables enter the outcome equation linearly, bias can still arise when the treatment variable \(T\) has a non-linear effect on the outcome. For instance, consider \(g(T,\vec{X},\upsilon) = f(T) + \vec{X}\gamma + \upsilon\). In this case, \(\beta\) would not, in general, equal the NRWE. To illustrate this, assume \(T \independent \upsilon \mid \vec{X}\) and \(E[\upsilon \mid \vec{X}] = 0\).Then
\[
\begin{aligned}
\cov(Y,\Delta(\vec{X}))
&= \cov\big(f(T) + \vec{X}\gamma + \upsilon,\;\Delta(\vec{X})\big) \\
&= \cov\big(f(T),\Delta(\vec{X})\big)
  + \cov\big(\vec{X}\gamma,\Delta(\vec{X})\big)
  + \cov\big(\upsilon,\Delta(\vec{X})\big) \\
&= \cov\big(f(T),\Delta(\vec{X})\big) \neq 0.
\end{aligned}
\]
The last equality uses (i) \(\cov(\vec{X},\Delta(\vec{X})) = 0\) by properties of the linear projection, implying
\(\cov(\vec{X}\gamma,\Delta(\vec{X})) = 0\), and (ii) \(E[\upsilon \mid \vec{X}] = 0\) implies \(\cov(\upsilon,\Delta(\vec{X})) = 0\).

This demonstrates that even when the conditional expectation function is linear in the control variables, misspecification bias can still arise if \(f(T)\) and \(\Delta(\vec{X})\) are correlated. For clear interpretability of the linear regression coefficients, it is important that the relationship between the explanatory variables themselves is linear, rather than necessarily requiring a linear relationship between the control variables and the outcome variable. Therefore, when selecting control variables for regression analysis, researchers should prioritize examining how the variable of interest interacts with the control variables. This consideration is more important than simply assessing the impact of the control variables on the outcome.

We can gain another perspective on the nature of misspecification bias by thinking of the misspecification error as the portion of variation that could potentially be explained but remains unexplained due to model constraints. Specifically, we can express \(\Delta(\vec{X})\) as the difference between residuals:

\[
\Delta(\vec{X}) = \mu(\vec{X}) - \pi \vec{X} = 
\underbrace{T - \pi \vec{X}}_{\parbox{4cm}{\footnotesize\centering Unexplained Due to \\ Linearity Restrictions}}  
- \underbrace{T - \mu(\vec{X})}_{\parbox{3.5cm}{\footnotesize\centering Fundamentally Unexplained}}.
\]

This equation shows that the bias arises from the variation in \(T\) that could potentially be explained by the control variables but remains unexplained due to the linearity assumptions of the model.\footnote{Note that \((T - \pi \vec{X}) - (T - \mu(\vec{X}))\)  is the residual from a linear projection of \(T - \pi \vec{X}\) onto \(T - \mu(\vec{X})\) .} If these unexplained components are correlated with the outcome variable, additional bias is introduced because the model fails to account for this variation. On the other hand, when there is no correlation, attenuation bias arises due to the misestimation of the effect of the control variables on \(T\), which stems from the linearity restriction.

Finally, while our focus is on the relationship between the data generating process and the coefficient \(\beta\), we can also express \(\beta\) as a weighted average of the local linear projection of \(Y\) on \(T\), along with a projection of \(Y\) on the misspecification error, as shown in Corollary \ref{cor1}.

\begin{corollary}\label{cor1}
Let \(\beta(\vec{X})\) denote the local regression coefficient of \(Y\) on \(T\), conditional on \(\vec{X} = \vec{x}\), and let \(\beta_{\Delta(\vec{X})}\) denote the regression coefficient from the population regression of \(Y\) on \(\Delta(\vec{X})\). Under the assumptions of Proposition \ref{yitzhakiGeneralizedThm}, the regression coefficient \(\beta\) in the population regression model \(Y = T \beta + \vec{X}\gamma + \epsilon\) is given by:

\[
\beta = E_{\vec{X}}[\beta(\vec{X}) \, w_1(\vec{X})] + \beta_{\Delta(\vec{X})} \, w_0,
\]
where
\[
w_1(\vec{X}) = \frac{\var(T \mid \vec{X})}{E_{\vec{X}}\left[\var(T \mid \vec{X}) + \var(\Delta(\vec{X}))\right]}, \quad 
w_0 = \frac{\var(\Delta(\vec{X}))}{E_{\vec{X}}\left[\var(T \mid \vec{X}) + \var(\Delta(\vec{X}))\right]}.
\]
\end{corollary}

Corollary \ref{cor1} demonstrates that the coefficient \(\beta\) in the multivariate regression can be interpreted as a weighted average of the local regression coefficients \(\beta(\vec{X})\), obtained by regressing \(Y\) on \(T\) within each level of \(\vec{X}\), along with a bias term \(\beta_{\Delta(\vec{X})} \, w_0\) arising from the misspecification error. The weights \(w_1(\vec{X})\) and \(w_0\) are determined by the variances of \(T\) conditional on \(\vec{X}\) and the variance of the misspecification error \(\Delta(\vec{X})\), respectively. This means that groups where \(T\) has more variability contribute more to the overall coefficient \(\beta\). 

\textit{Under what conditions can we expect that the coefficient on the variable of interest captures the $NRWE$?} First, if our regression is fully saturated and all control variables are discrete, then the linear approximation of the conditional expectation is exact, and the misspecification error is zero, $\Delta(x) = 0$ for all $x$, eliminating both the bias term and the attenuation effect. Another example where the bias is zero occurs when the joint distribution of the explanatory variable (and not necessarily the joint distribution of the outcome variables and explanatory variables) belongs to the Elliptically Contoured distributions\footnote{The Elliptically Contoured distributions famously include the multivariate Gaussian distribution.}; here, the conditional expectation of the variable of interest is linear in the other variables, avoiding the misspecification bias and attenuation bias. Additionally, in some instances, including sufficient interaction terms between variables can approximate the underlying data-generating function, thereby reducing biases (e.g., \cite{hastie2009elements}).

In general, the linearity assumption is unlikely to hold, and both misspecification bias and attenuation bias may arise, necessitating a thorough evaluation of the relationships between variables. For instance, if higher values of the control variables $\text{X}$ tend to increase both the variable of interest and the outcome variable in a convex manner, then the linear projection will likely underestimate $T$ at high values of $X$. This implies that $\Delta(X)$ is likely to be positive for higher values of $X$, and consequently, it may be positively correlated with the outcome $Y$, suggesting $\cov\left(Y,\Delta(X)\right)\geq 0$. Such a scenarios would induce an upward bias in $\beta$ compared to the weighted effect component. 

For example, consider researchers exploring the effect of parent income on a child's years of schooling while controlling for the parent level of education. Previous studies have shown that average income grows exponentially with years of schooling\footnote{Usually the relation is described as log-linear.} (e.g., \cite{mincer1974schooling}, \cite{heckman2003fifty}). Hence, the relationship between parent schooling and income is likely to be increasing, and linear projection would underestimate parent income at high values. Since the parent years of schooling are likely to be positively correlated with a child's years of schooling, the estimate for the effect of parent income is likely to be biased. 

On the other hand, if the researchers want to flip prespective, and examine the influence of parent educational level on a child's educational attainment while controlling for income, the role of the control variable and the variable of interest is reversed. In this case, the average parent years of schooling is a concave function (log) of parent income. Then, $\Delta(X)$ will likely be negative for higher values of $X$ (linear projection overestimates $T$ at high values $X$). In this case, if parent income increases a child's years of schooling in a convex manner, then the misspecification bias is likely to be negative.

\subsection{Why the NRWE Is of Interest and Why It Is the Baseline Estimand Against Which We Measure Bias}
In this section, we discuss three key properties of the NRWE estimand that make it interesting in its own right and a useful baseline against which to measure regression-coefficient bias. The first appealing property of the NRWE is its ease of computation. When the relationship between T and X is linear, the NRWE coincides with the OLS coefficient, so it can be obtained directly from a standard regression. This ease of computation is particularly attractive: under a simple linearity condition on the relationship between T and X, the NRWE requires no nonparametric smoothing or derivative estimation. It is exactly the coefficient that applied researchers already compute using OLS.

Second, from a reverse perspective, the NRWE weights offer a unique way to express the OLS coefficient $\beta$ as a convex weighted average of the changes in the conditional expectation function. Specifically, for a given distribution of $(T, X)$, there is essentially a single set of weights that represents $\beta$ as an average of the partial derivatives $\partial_t E[Y|t,x]$ across all admissible data generating processes. As discussed above, this is appealing because the usual interpretation of $\beta$ is that it captures how a change in $T$ affects the expected value of $Y$, holding the controls variables fixed. The weighted interpretation of the $\beta$ coefficient provides a concrete and explicit formalization of this idea. We establish this notion of uniqueness in Proposition~\ref{prop:nrwe_uniqueness_global} below.

\begin{prop}[Uniqueness the NRWE weights]\label{prop:nrwe_uniqueness_global}
Fix a joint distribution of $(T,X)$. Assume that for almost every $x\in\mathcal X$, the conditional distribution of $T$ given $X=x$ admits a density $f_{T\mid X}(\cdot\mid x)$ with finite, strictly positive variance $0<\Var(T\mid X=x)<\infty$ and $E[Y|t,x]<\infty$ for every $t$ and $x$. Consider an arbitrary data-generating processes of the form
\[
Y = m(T,X) + \varepsilon, 
\qquad E[\varepsilon\mid T,X]=0,
\]
where, for each $x$, the function $t\mapsto m(t,x)$ is differentiable with continuous derivative $\partial_t m(t,x)$ and satisfy $E[|\partial_t m(T,X)|\big] < \infty$. For each such data-generating process, let $\beta$ denote the coefficient on $T$ in the population OLS regression of $Y$ on $T$ and $X$. Assume that $E[T|X]$ is a linear function such that, by proposition \ref{yitzhakiGeneralizedThm}, $\beta$ equals to the NRWE. Finally, denote the full NRWE weights as $b^\star(t,x) = f_X(x)w(t,x)$, where $f_X(x)$ denote the density of $x$ and $w(t,x)$ are the within cell NRWE weights defined above.   

Now suppose there exists another weight function $a:\mathbb R\times\mathcal X\to\mathbb R$ such that:
\begin{enumerate}
\item For every data-generating process of the above form with this fixed $(T,X)$,
\begin{equation}
\label{eq:global_alt_rep}
\beta
= \int \partial_t m(t,x)\,a(t,x)\,dt\,dx.
\end{equation}
\item For each $x$, the map $t\mapsto a(t,x)$ is continuous and differentiable on $\mathbb R$, and
\[
\int\!\!\int \big|a(t,x)\big|\,dt\,dx < \infty.
\]
\end{enumerate}
Then $a(t,x)=b^\star(t,x)$ for almost every $(t,x)$.
\end{prop}
Proof is in the Appendix. The proposition shows that, once the joint distribution of $(T, X)$ is fixed, the weights $b^\star(t, x) = f_X(x) w(t, x)$ are the unique measurable function of $(t, x)$ that expresses the regression coefficient $\beta$ as an average of $\partial_t m(t, x)$ over all structural functions $m$. These results highlight that the NRWE weights provide the only interpretation of $\beta$ as a weighted average of $\partial_t E[Y \mid t, x]$, making explicit the idea that $\beta$ captures the expected change in the conditional mean of $Y$ when $T$ increases while $X$ is held fixed.

A third appealing property of the NRWE is its efficiency. It provides the weighting scheme for
$\partial_t E[Y \mid t, x]$
that attains the lowest semiparametric efficiency bound within a broad admissible class. Proposition \ref{prop:efficiency_global} shows that, under homoskedasticity, the NRWE weights uniquely minimize this bound among all weights that integrate to one and satisfy a boundary condition.

\begin{prop}[Semiparametric Efficiency of the NRWE Weights]
\label{prop:efficiency_global}

Let $(Y_i,T_i,X_i)_{i=1}^n$ be an i.i.d.\ sample. Assume
\begin{align*}
Y &= m(T,X) + \varepsilon, 
\qquad E[\varepsilon\mid T,X]=0,
\qquad \Var(\varepsilon\mid T,X)=\sigma^2\in(0,\infty),\\
f_{T\mid X}(t\mid x) &>0 \text{ for all } t\in \mathrm{supp}(T\mid X=x), \qquad E_X[\Var(T\mid X)]\in(0,\infty).
\end{align*}
Let $f_{T,X}(t,x)$ and $f_X(x)$ denote the joint and marginal densities of $(T,X)$ and $X$,
respectively, so that $f_{T,X}(t,x)=f_{T\mid X}(t\mid x)f_X(x)$. Let $a(t,x)$ be a weighting function that is absolutely continuous in $t$ for each $x$, and define the
\emph{globally weighted average derivative}
\[
\theta_a
\;=\;
\int \partial_t m(t,x)\,a(t,x)\,dt\,dx.
\]

We restrict attention to weights $a$ satisfying the following:

\begin{enumerate}[label=(R\arabic*)]
\item \textbf{(Boundary)} For each $x$, $a(\cdot,x)$ is absolutely continuous in $t$ and satisfies
\[
\lim_{t\to \partial \mathrm{supp}(T\mid X=x)} a(t,x)=0,
\qquad
\lim_{t\to \partial \mathrm{supp}(T\mid X=x)} t\,a(t,x)=0,
\]
if the support is unbounded,
interpret as $t\to\pm\infty$.
\item \textbf{(Integrate to 1)}
\[
\int a(t,x)\,dt\,dx = 1.
\]
\end{enumerate}

Under regularity Assumption \ref{assump:efficiency_global} listed in the Appendix, among all weighting functions $a$ satisfying \textup{(R1)} and \textup{(R2)}, the semiparametric efficiency bound for estimating $\theta_a$ under homoskedasticity is minimized at 
\[
a^\star(t,x)
=
\frac{\displaystyle \int_t^\infty (u-\mu(x))\, f_{T|X}(u|x)f_X(x)\,du}
     {\displaystyle E_X[\Var(T\mid X)]}.
\]
which are exactly the NRWE weights. The corresponding minimum asymptotic variance is
\[
\mathcal{V}_{\min}
=\frac{\sigma^2}{E_X[\Var(T\mid X)]}.
\].
\end{prop}
The proof is in the Appendix, and follows standard argument for semiparametric efficiency bounds (\cite{vanderVaart1998},\cite{Kennedy2024SDRTDML},\cite{NeweyStoker1993}). The proposition provides an efficiency-based justification for the NRWE weights. For any weighting function \(a(t,x)\) satisfying \textup{(R1)}–\textup{(R2)}, the corresponding functional \(\theta_a\) has a semiparametric efficiency bound under homoskedasticity. Proposition~\ref{prop:efficiency_global} shows that this bound is minimized exactly at the NRWE weights, and that the minimum asymptotic variance is \(\sigma^2 / E_X[\Var(T\mid X)]\). In other words, among all normalized and well-behaved weighting schemes in this class, the NRWE target is the statistically “cheapest” globally weighted average derivative to estimate.

The restrictions \textup{(R1)}–\textup{(R2)} themselves are mild. The normalization \(\int a(t,x)\,dt\,dx=1\) ensures that \(\theta_a\) is a genuine average of marginal effects rather than an arbitrarily scaled contrast. The boundary condition in \textup{(R1)} rules out weights that load heavily on the extremes of \(\mathrm{supp}(T\mid X=x)\), precisely where \(\partial_t m(t,x)\) is typically poorly identified in continuous-treatment settings because those treatment levels are rarely observed. By keeping the weights small near the boundaries, \textup{(R1)} keeps \(\theta_a\) within the convex hull of locally identified effects and prevents the estimand from being driven by extrapolation in the tails. The admissible class is therefore quite large and flexible, but it excludes exactly those functionals that would be most sensitive to ill-posedness at the edges of the support.

In causal applications, closely related restrictions are often motivated by sign preservation. \citet{blandhol2022tsls}, for example, argue that linear IV estimands should satisfy “weak causality”: if all underlying causal effects of shifts from \(t\) to \(t'>t\) are nonnegative, then the resulting scalar estimand should also be nonnegative. A convenient way to guarantee this property is to average causal effects using nonnegative weights that integrate to one. Although Proposition \ref{prop:efficiency_global} does not impose nonnegativity, the NRWE weights are in-fact nonnegative, so that \(\beta\) can be interpreted as a sign-preserving average of marginal causal effects.

Taken together, these features explain why the NRWE is a natural baseline estimand for assessing misspecification biases. It is directly estimable by OLS when the relationship between \(T\) and \(X\) is linear, it is the unique derivative-based representation of the regression coefficient for a fixed design, and it is asymptotically efficient within a broad class of average-derivative estimands. When researchers seek a single scalar summary of the heterogeneous marginal effect \(\partial_t m(T,X)\), the NRWE combines attractive properties with a transparent causal interpretation.

\begin{remark}
A natural question is why the standard interpretation of the regression coefficients as the best linear approximation to the conditional expectation function is not already sufficient for interpretation of $\beta$. While this perspective can be useful in some cases, it is not tied to how applied work tends to reason about regression and can be misleading for interpretation. Researchers typically read \(\beta\) as summarizing how the conditional mean of \(Y\) changes when \(T\), the variable of interest, \emph{changes}, holding \(X\) fixed. However, a best linear approximation to the level of the conditional expectation need not nessarly to approximate its derivative well; the linear predictor can fit the function in a least-squares sense while still providing a poor summary of local changes in \(T\). In addition, “best” is defined relative to the empirical joint distribution of \((T,X)\), so the approximation is implicitly weighted toward regions of the covariate space where the regressors are more frequently observed, or the prediction error is larger. The NRWE formulation makes this dependence on the distribution of \((T,X)\) explicit, whereas the best-linear-approximation language leaves these weights implicit. Finally, the approximation is taken in the full regressor space \((T,X)\), not with respect to the effect of \(T\) alone, so a-priori, there is no guarantee that the linear predictor is especially accurate along the “direction” of the variable of interest; it may fit well in some combinations of regressors while still providing a distorted summary of how \(Y\) responds to variation in \(T\). For these reasons, it is more transparent to interpret \(\beta\) through the NRWE and to treat any discrepancy as a misspecification term rather than relying solely on the best-linear-approximation interpretation. 
\end{remark}

\subsection{Numerical Illustration}
In this section, we demonstrate, using a numerical example, that under different data generating process, the size of two biases, the attenuation and misspecification bias, can be substantial and lead to incorrect conclusions. We assume the following DGP:
\be
T = h(X) + \nu \\ 
Y = g(T,X) + \epsilon,
\ee
where $h(\cdot)$ and $g(\cdot)$ will be defined later and  
$$
X \sim U(0,5),\quad \nu \sim N(0,\sigma),\quad \epsilon \sim N(0,1).
$$
In this DGP, where the conditional distribution of $T|X$ is Normal, we can derive a closed form expression for the weights derived in Proposition \ref{yitzhakiGeneralizedThm}. Specifically, we have the following:
\be
& E[T - E[T|X] \mid T > t, X] \cdot P(X > t \mid X) = \\ & \left[h(X) + \sigma \frac{\phi\left(\frac{t-h(X)}{\sigma}\right)}{1 - \Phi\left(\frac{t-h(X)}{\sigma}\right)} - h(X)\right] \cdot \left(1 - \Phi\left(\frac{t-h(X)}{\sigma}\right)\right)   = \\ 
&  \sigma \phi\left(\frac{t-h(X)}{\sigma}\right).
\ee
Similarly, the NRWE weights normalize by $E[\Var(T\mid X)]$. Under the DGP $T=h(X)+\nu$ with homoskedastic $\nu\sim N(0,\sigma^2)$, we have $\Var(T\mid X)=\Var(\nu\mid X)=\sigma^2$ for all $X$, hence $E[\Var(T\mid X)]=\sigma^2$.
\[
w(t, X) = \frac{\sigma \phi\left(\frac{t-h(X)}{\sigma}\right)}{\sigma^2} = \frac{\phi\left(\frac{t-h(X)}{\sigma}\right)}{\sigma}
\]
which is simply the Normal distribution density function, and we can approximate $NRWE$ numerically by generating samples and estimating the mean derivative in the population 
$$
\E\left[\frac{\partial g(t,x)}{\partial t} \right] \approx \frac{1}{n} \sum_{i=1}^n \frac{\partial g(t_i,x_i)}{\partial t_i} .
$$
where $n$ is the number of simulated samples. Similarly, we can approximate the bias term by calculating the sample covariance, $\hat{\cov}(Y,h(X)-\pi_T\vec{X})$, and the sample variance of the residualized $T$, where $\hat{\pi}_{T}\vec{X}$ is the regression coefficient on $\vec{X}$ in a linear regression of $T$ on $\vec{X}$.

In our simulation, we generate 1,000,000 draws and run a Monte Carlo simulation with 300 iterations, setting \(\sigma = 1\). The results of these simulations, shown in Table \ref{simTalbe}, explore different data generating processes.

The first row illustrates the point we made in Section \ref{NRWEDiscussion}. In this case, the relationship between the control variable \(X\) and the variable of interest \(T\) is convex, and higher values of \(X\) are associated with higher outcome values. Since the linear model underestimates the true value for large \(X\), the estimated coefficient \(\beta\) is much larger than the NRWE, driven by misspecification bias. Additionally, the attenuation bias—the difference between the weighted effect of \(T\) and the NRWE—is substantial, with the weight on the marginal expectation derivatives approaching zero.

The second row illustrates that linearity in the variable of interest is not, by itself, enough to guarantee an unbiased regression coefficient. Although the NRWE equals $1$, the estimated coefficient is $1.997$. This discrepancy reflects two distinct components in the decomposition: the weighted-effect component is almost fully attenuated ($\approx 0.003$), creating a large negative gap relative to NRWE (about $-1$), while the misspecification term is large and positive ($\approx1.994$). Thus, the coefficient’s excess over the attenuated weighted effect is entirely due to misspecification, whereas the gap between NRWE and the weighted effect reflects attenuation.

The third row highlights how regression coefficients can lead to incorrect conclusions about the relationship between \(Y\) and \(T\). In this case, even though \(T\) has no direct effect on the outcome variable, the regression yields a significant coefficient. This coefficient is entirely driven by misspecification bias, demonstrating that misleading results can arise when the model is misspecified.

\begingroup
\singlespacing
\begin{table}[H]
    \centering
    \footnotesize
    \resizebox{\textwidth}{!}{%
        \begin{tabular}{@{}lccccc@{}}
\toprule
 & \multicolumn{1}{l}{\textbf{NRWE}} & \multicolumn{1}{l}{\textbf{$\beta$}} & \multicolumn{1}{l}{\textbf{Misspecification Bias}} & \multicolumn{1}{l}{\textbf{Weighted Effect of T}} & \multicolumn{1}{l}{\textbf{Attenuation Bias}} \\
\midrule
\begin{tabular}[c]{@{}l@{}}
$\mathbb{E}[T\mid X] = \exp(X)$\\
$\mathbb{E}[Y\mid T,X] = \sin^2(T) + X$
\end{tabular} & -0.0414 & 0.0049 & 0.0051 & -0.0001 & -0.0413 \\
 & (0.0007) & (0.0001) & (0.0001) & (0.0000) & (0.002) \\
\begin{tabular}[c]{@{}l@{}}
$\mathbb{E}[T\mid X] = \exp(X)$\\
$\mathbb{E}[Y\mid T,X] = T + \exp(X)$
\end{tabular} & 1.000 & 1.997 & 1.994 & 0.003 & 0.995 \\
 & NA & (0.0001) & (0.0002) & (0.0002) & (0.0733) \\
\begin{tabular}[c]{@{}l@{}}
$\mathbb{E}[T\mid X] = \sin(X)$\\
$\mathbb{E}[Y\mid T,X] = \sin^2(X) + X^2$
\end{tabular} & 0.000 & -0.512 & -0.512 & 0.000 & 0.000 \\
 & NA & (0.0018) & (0.0060) & (0.0062) & (0.0010) \\
\bottomrule
\end{tabular}

    }
    \caption{Simulation Results}\label{simTalbe}
    \justifying
    \small\textit{Notes: This table presents the results from a Monte Carlo exercise that calculates the decomposition of the regression coefficient $\beta$, according to Proposition \ref{yitzhakiGeneralizedThm}, from the regression model $Y = \beta T + \alpha X + u$, where the data generating process is specified in the first column. The coefficient $\beta$ is decomposed into the misspecification bias and the weighted effect of $T$. The last column shows the attenuation bias, calculated as the difference between the Naïve Regression Weighted Effect and the weighted effect of $T$. Standard deviations of the estimated parameters are in parentheses.}
\end{table}
\endgroup

\section{Conclusion}
Proposition \ref{yitzhakiGeneralizedThm} emphasizes the difficulties in interpreting regression coefficients when the underlying data-generating process is not linear. However, it also provides guidance on how researchers can address these biases when interested in the $NRWE$ parameter. The simplest approach to obtain an unbiased estimate of $NRWE$ is to include an estimate of $E[T|\vec{X}]$ as a control variable in the regression\footnote{This is because the conditional expectation, given the conditional expectation, is a trivial linear function of the conditional expectation: $E[T|E[T|X]] = E[T|X]$}. To compute $E[T|\vec{X}]$, one can either estimate it nonparametrically (e.g., \cite{ullah1999nonparametric}), or use prior knowledge of it (\cite{borusyak2021nonrandom}). However, this raises a question: if one can estimate $E[T|\vec{X}]$ nonparametrically, why choose to estimate the causal effect using regression instead of estimating the entire model nonparametrically? In many cases, researchers opt for linear regression due to its efficiency and stability, two properties that do not always characterize nonparametric estimators, especially when the dimensions of $\vec{X}$ are large. Therefore, if researchers wish to use regression, it would be insightful to include in their analysis a discussion on the relationship between the control variables and the variable of interest. This can be done, for example, by plotting $E[T|x_j]$ for different components of $\vec{X}$, or provide theoretical justification for the use of linear controls. 

Researchers should also bear in mind that the relationship between the variables of interest and control variables are not generally invariant to monotonic changes. For instance, researchers should be cautious when estimating a linear model where $T$ enters the regression linearly, and a similar model where they use $\log(T)$ instead. Without altering the control variable as well, as Proposition \ref{yitzhakiGeneralizedThm} shows, both models are unlikely to obtain a weighted average of changes in the conditional expectation, and at least one of them is likely to suffer from a misspecification bias. Hence, researchers should be more conscious of how they model their control variables.
\bibliographystyle{plainnat} 
\bibliography{bibFile} 
\newpage
\appendix
\section{Appendix}
\subsection{Proof of Proposition 1}
Denote by $\pi$ the coefficients from linear projection of $T$ on $\vec{X}$. Denote by $f(T)$ and $f(T|\vec{X})$ the density and conditional density of $T$. Using Frisch-Waugh-Lovell theorem (\cite{frisch1933partial}), we have that 
\be
\beta = \frac{\cov(Y,T - \vec{X}\pi)}{\var(T - \pi \vec{X})}.
\ee
We start by focusing the numerator. Denote the conditional expectation of $T$, condition on $\vec{X}$ by $\mu(\vec{X}) $. Then we can express the numerator as 
\be
\cov(Y,T - \vec{X}\pi) & = \cov(Y,(T - \mu(\vec{X}) + \mu(\vec{X}) -\vec{X}\pi) \\
& = \cov(Y,(T - \mu(\vec{X})) + \cov(Y,\mu(\vec{X}) -\vec{X}\pi).
\ee
Using the law of iterated expectations and integration by parts we can re-express the first term as 
\begin{align*}
\operatorname{Cov}\bigl(Y, T - \mu(\vec X)\bigr)
&= \mathbb E\bigl[ Y (T - \mu(\vec X)) \bigr] \\
&= \mathbb E_{\vec X}\Bigl[ \mathbb E\bigl[ Y (T - \mu(\vec X)) \mid \vec X \bigr] \Bigr] \\
&= \mathbb E_{\vec X}\Bigl[ \mathbb E\bigl[ \mathbb E[Y \mid T,\vec X] (T - \mu(\vec X)) \mid \vec X \bigr] \Bigr] \\
&= \mathbb E_{\vec X}\biggl[
    \int_{-\infty}^{\infty}
        \mathbb E[Y \mid T=t,\vec X]\,
        (t - \mu(\vec X))\,
        f_{T \mid \vec X}(t \mid \vec X)\, dt
\biggr] \\
&= \mathbb E_{\vec X}\biggl[
    \int_{-\infty}^{\infty}
        \frac{\partial}{\partial t} \mathbb E[Y \mid T=t,\vec X]\,
        \biggl( - \int_{-\infty}^{t}
            (u - \mu(\vec X))\, f_{T \mid \vec X}(u \mid \vec X)\, du
        \biggr)\, dt
\biggr] \\
&= \mathbb E_{\vec X}\biggl[
    \int_{-\infty}^{\infty}
        \frac{\partial}{\partial t} \mathbb E[Y \mid T=t,\vec X]\,
        \mathbb E[T - \mu(\vec X) \mid T>t,\vec X]\,
        \Pr(T>t \mid \vec X)\, dt
\biggr].
\end{align*}
Where the last equality follows from the fact that 
\be
E[T-\mu(\vec{X})| T>t, \vec{X}]\pr(T>t| \vec{X}) + 
E[T-\mu(\vec{X})| T\leq t, \vec{X}]\pr(T\leq t| \vec{X}) = 0. 
\ee
Therefore the numerator is given by 
\be
\cov(Y,T - \vec{X}\pi) & = E_{\vec{X}} \bigg[   \int_{-\infty}^{\infty} \frac{\partial E[Y|t,\vec{X}]}{\partial t } E[T-\mu(\vec{X})|T>t,\vec{X}]p(T>t|\vec{X})dt \bigg] + \cov(Y,\mu(\vec{X}-\pi \vec{X}).
\ee

Next, we turn to the denominator. We can re-express it as 
\be
\var(T - \pi \vec{X}) & = \var(
T - \mu(\vec{X}) + \mu(\vec{X}) -  \pi \vec{X})  \\
& = \var(T - \mu(\vec{X})) + \var(\mu(\vec{X}) -  \pi \vec{X}) ) + 2\cov(T - \mu(\vec{X}), \mu(\vec{X}) -  \pi \vec{X} ) \\
& = \var(E[T-\mu(\vec{X})|\vec{X}]) + E[\var(T|\vec{X})] + \var(\mu(\vec{X}) -  \pi \vec{X}) ) \\
& = \E[\var(T|\vec{X})] + \var(\mu(\vec{X}) -  \pi \vec{X}) ),
\ee
where we used the law of total variance and the fact that $\cov(T-\mu(\vec{X}),\mu(\vec{X})-\pi \vec{X}) = E[(T-\mu(\vec{X})(\mu(\vec{X})-\pi \vec{X}))] = 0$, due to the law of iterated expectations, which concludes the proof. 

\subsection{Proof of Corollary \ref{cor1}}
By the results of Proposition \ref{yitzhakiGeneralizedThm}, we can express
\be
\beta 
&= \mathbb{E}_{\vec{X}}\bigg[ \int_{-\infty}^{\infty} 
    \frac{\partial \mathbb{E}[Y \mid T=t,\vec{X}]}{\partial t } 
    \, w(t,\vec{X}) \, dt \bigg] 
  + \frac{\cov(Y,\Delta(\vec{X}))}
    {\mathbb{E}_{\vec{X}}[\var(T \mid \vec{X})] + \var(\Delta(\vec{X})) } \\
&= \mathbb{E}_{\vec{X}}\bigg[ \int_{-\infty}^{\infty} 
    \frac{\partial \mathbb{E}[Y \mid T=t,\vec{X}]}{\partial t } 
    \, w(t,\vec{X}) \frac{\var(T \mid \vec{X})}{\var(T \mid \vec{X})} \, dt \bigg] 
  + \frac{\cov(Y,\Delta(\vec{X}))}
    {\mathbb{E}_{\vec{X}}[\var(T \mid \vec{X})] + \var(\Delta(\vec{X})) }
    \frac{\var(\Delta(\vec{X}))}{\var(\Delta(\vec{X}))}.
\ee

Recall that the weight function can be written as
\[
w(t,\vec{X}) 
= w_{\text{Yitzhaki}}(t,\vec{X}) \,
  \frac{\var(T \mid \vec{X})}{\mathbb{E}_{\vec{X}}[\var(T \mid \vec{X})] 
    + \var(\Delta(\vec{X}))},
\]
where $w_{\text{Yitzhaki}}$ are the univariate weights, 
as stated in equation~\eqref{yithzhakiResult}.  Substituting this expression
for $w(t,\vec{X})$ and using the definitions of 
$\beta(\vec{X})$, $w_{1}(\vec{X})$, $\beta_{\Delta(\vec{X})}$, and $w_{0}$ gives
\be
\beta 
&= \mathbb{E}_{\vec{X}}\big[ \beta(\vec{X}) w_{1}(\vec{X}) \big] 
  + \beta_{\Delta(\vec{X})} \, w_{0}.
\ee

In the second line we multiply numerator and denominator by the corresponding
variance terms, and in the last line we apply the identity above
(equation~\eqref{yithzhakiResult}) together with the definitions of the
local coefficients and weights.

\subsection{Proof of Proposition \ref{prop:nrwe_uniqueness_global}: Uniqueness of the Weights}

\begin{proof}
We want to show that $a(t,x)=b^\star(t,x)$ almost everywhere for this fixed distribution of $(T,X)$. Define 
\[
H(t,x) := a(t,x) - b^\star(t,x)
\]
Since by assumption both $a$ and $b^\star$ are absolutely integrable, their difference
is also absolutely integrable. Next, Subtracting 
\begin{equation}
\label{eq:global_nrwe_rep}
\beta
= \int \partial_t m(t,x)\,b^\star(t,x)\,dt\,dx,
\end{equation}
and 
\begin{equation}
\label{eq:global_alt_rep}
\beta
= \int \partial_t m(t,x)\,a(t,x)\,dt\,dx,
\end{equation}
gives us
\begin{equation}
\label{eq:H_orth_global}
\int \partial_t m(t,x)\,H(t,x)\,dt\,dx
= 0
\qquad\text{for all admissible }m.
\end{equation}
Now, as this should hold for every $m$, we can examine a specific $m$. fix any $\phi\in C_c^\infty(\mathbb R)$ and any bounded measurable $g:\mathcal X\to\mathbb R$.
Define
\[
h(t):=\int_{-\infty}^t \phi(s)\,ds,
\qquad
m(t,x):=g(x)\,h(t).
\]
Then $m$ is admissible since
$E\big[|\partial_t m(T,X)|\big]=E\big[|g(X)\phi(T)|\big]\le \|g\|_\infty E[|\phi(T)|]<\infty$,
and $\partial_t m(t,x)=g(x)\phi(t)$.
Plugging this into \eqref{eq:H_orth_global} yields
\[
\iint g(x)\phi(t)H(t,x)\,dt\,dx=0.
\]
By Fubini's theorem,
\[
\int g(x)\,F_\phi(x)\,dx=0\quad\text{for all bounded measurable }g,
\qquad
F_\phi(x):=\int \phi(t)H(t,x)\,dt.
\]
Hence $F_\phi(x)=0$ for almost every $x$. Therefore, for almost every $x$,
\[
\int \phi(t)H(t,x)\,dt=0\quad\text{for all }\phi\in C_c^\infty(\mathbb R).
\]
This implies $H(\cdot,x)=0$ almost everywhere in $t$. Consequently, $H(t,x)=0$ for almost every $(t,x)$,
so $a(t,x)=b^\star(t,x)$ almost everywhere.

\end{proof}

\newpage

\subsection{Proof of Proposition \ref{prop:efficiency_global}: Efficiency}

We start by introducing a set of regularity conditions 
\begin{assumption}
\label{assump:efficiency_global}
The following conditions hold:
\begin{enumerate}
\item[(A1)] $(Y,T,X)$ are i.i.d.\ with finite second moments: $E[Y^2],E[T^2]<\infty$.
\item[(A2)] The conditional density $f_{T\mid X}(t\mid x)$ is continuously differentiable and bounded away
from $0$ on its support, and $f_X(x)$ is bounded on its support.
\item[(A3)] The weight $a(t,x)$ satisfies \textup{(R1)} and \textup{(R2)} above, and $m(\cdot,x)$ is
absolutely continuous in $t$, bounded $|m(t,x)| < \infty$ for all $x,t$, and with $\partial_t m(t,x)$ integrable against all admissible weights:
\[
\int\!\!\int \big|\partial_t m(t,x)\,a(t,x)\big|\,dt\,dx < \infty.
\]
Moreover, for each $x$, $\partial_t m(\cdot,x)$ is integrable against all admissible $a(\cdot,x)$.
\item[(A4)] The conditional law $f_{T\mid X}$ (equivalently, $f_{T,X}$) is known (e.g.\ design-based
ignorability) or estimable at $n^{1/2}$-rate.
% \item[(A5)] Homoskedasticity: $\Var(\varepsilon\mid T,X)=\sigma^2\in(0,\infty)$.
\end{enumerate}
\end{assumption}
\noindent We can now proceed with the proof.

\begin{proof}
Throughout, let $a(t,x)$ be absolutely continuous in $t$ for each $x$ and satisfy the
\textbf{regularity / boundary} and \textbf{global mass-one} conditions:
\[
\tag{R}\lim_{t\to \partial \mathrm{supp}(T\mid X=x)} a(t,x)=0,
\qquad
\lim_{t\to \partial \mathrm{supp}(T\mid X=x)} t\,a(t,x)=0,
\]
and
\[
\tag{N}\int a(t,x)\,dt\,dx=1.
\]
Define
\[
k(t,x):=-\partial_t a(t,x).
\]

For each fixed $x$, integration by parts in $t$ gives
\[
\int \partial_t m(t,x)\,a(t,x)\,dt
=\Big[m(t,x)a(t,x)\Big]_{\partial \mathrm{supp}}-\int m(t,x)\,\partial_t a(t,x)\,dt.
\]
Under (R1) and boundness of $m(t,x)$, the boundary term vanishes and we obtain
\[
\int \partial_t m(t,x)\,a(t,x)\,dt
=\int m(t,x)\,k(t,x)\,dt,
\]
so that
\[
\theta_a
=\int\!\!\int m(t,x)\,k(t,x)\,dt\,dx.
\]

The boundary condition (R1) implies two moment restrictions on $k$. First, for each $x$,
\[
\int k(t,x)\,dt
=\int -\partial_t a(t,x)\,dt
=-a(t,x)\Big|_{\partial \mathrm{supp}}=0.
\]
Second,
\begin{align*}
\int\!\!\int t\,k(t,x)\,dt\,dx
&=\int\!\left[\int t(-\partial_t a(t,x))\,dt\right]dx\\
&=\int\!\left[-t a(t,x)\Big|_{\partial \mathrm{supp}}+\int a(t,x)\,dt\right]dx
=\int\!\!\int a(t,x)\,dt\,dx
=1,
\end{align*}
where we used $\lim_{t\to\partial \mathrm{supp}} t a(t,x)=0$ and then (R2). Thus (R1) and (R2) imply 
\begin{equation}
\label{eq:k_constraints_global}
\int k(t,x)\,dt=0\ \ \forall x,
\qquad
\int\!\!\int t\,k(t,x)\,dt\,dx=1.
\end{equation}
Conversely, given any $k$ satisfying \eqref{eq:k_constraints_global} and suitable integrability,
\[
\int |k(t,x)|dt < \infty, \quad \int |t k(t,x)| < \infty, 
\]
the construction
\[
a(t,x):=\int_t^\infty k(u,x)\,du
\]
produces an absolutely continuous $a$ that satisfies (R1) and (R2):
\begin{itemize}
\item $a(\cdot,x)$ is absolutely continuous with $-\partial_t a(t,x)=k(t,x)$.
\item As $t\to\infty$, $a(t,x)\to 0$ by definition; as $t\to -\infty$ (or to the lower endpoint of
the support), $a(t,x)\to \int k(u,x)\,du=0$ by the first constraint in
\eqref{eq:k_constraints_global}, so $a$ satisfies (R).
\item The second constraint in \eqref{eq:k_constraints_global} then ensures
\[
\int\!\!\int a(t,x)\,dt\,dx
=\int\!\!\int t\,k(t,x)\,dt\,dx
=1.
\]
\end{itemize}
Hence minimizing over admissible $a$ is equivalent to minimizing over $k$ satisfying
\eqref{eq:k_constraints_global}, and we may work directly with $k$.

The following steps derive the semiparametric efficiency bound and build on standard arguments (e.g \cite{vanderVaart1998} \cite{Kennedy2024SDRTDML},\cite{NeweyStoker1993}). Let $\mathcal P$ be the nonparametric conditional-mean model imposing only
$E[\varepsilon\mid T,X]=0$ and homoskedasticity $\Var(\varepsilon\mid T,X)=\sigma^2$. Write $\varepsilon := Y - m(T,X)$. Consider the Gaussian regression submodel with fixed \(f_{T,X}\):
\[
 Y\mid T=t,X=x \sim \mathcal N(m_\eta(t,x),\sigma_\eta^2),
\quad m_\eta=m+\eta h,\quad \sigma_\eta^2=\sigma^2+\eta b.
\]
Then the scores at \(\eta=0\) are
\[
 s_h=\frac{\varepsilon}{\sigma^2}h(T,X),\qquad
 s_b=b\cdot\frac{\varepsilon^2-\sigma^2}{2\sigma^4}.
\]
 
Using the representation $\theta_a=\int\!\!\int m(t,x)k(t,x)\,dt\,dx$, the pathwise
derivative in direction $h$ is
\begin{equation}
\label{eq:path_derivative_global}
D\theta_a[h]
=\int\!\!\int h(t,x)\,k(t,x)\,dt\,dx.
\end{equation}
which can be written as
\[
\int\!\!\int h(t,x)\,k(t,x)\,dt\,dx
=E\!\left[\frac{k(T,X)}{f_{T,X}(T,X)}\,h(T,X)\right],
\]
whenever $k/f_{T,X}$ is square-integrable. The efficient influence function $\phi_a$ must satisfy
$E[\phi_a\,s_h]=D\theta_a[h]$ for all square-integrable $h$.
That is,
\[
E\!\left[\phi_a(Y,T,X)\frac{Y-m(T,X)}{\sigma^2}\,h(T,X)\right]
=
E\!\left[\frac{k(T,X)}{f_{T,X}(T,X)}\,h(T,X)\right]
\quad\forall h.
\]
By the Riesz representation theorem in $L^2(P_{T,X})$, this yields 
\[
\phi_a(Y,T,X)
=\frac{k(T,X)}{f_{T,X}(T,X)}\,(Y-m(T,X)),
\]
up to addition of mean-zero functions orthogonal to the tangent space, which do not reduce variance. 

Write $w(T,X) := k(T,X)/f_{T,X}(T,X)$, so that
\[
\phi_a(Y,T,X) = w(T,X)\,\varepsilon.
\]
Since $\theta_a$ does not depend on $\sigma^2$, the pathwise derivative in any variance direction $b$ is zero, so the efficient influence function must also satisfy $E[\phi_a s_b]=0$ for all $b$. Using the expression for $s_b$,
\[
E[\phi_a s_b]
= \frac{b}{2\sigma^4} E\big[w(T,X)\,\varepsilon(\varepsilon^2-\sigma^2)\big]
= \frac{b}{2\sigma^4} E\big[w(T,X)\,E(\varepsilon^3\mid T,X)\big].
\]
Under the Gaussian submodel, $\varepsilon\mid T,X \sim \mathcal N(0,\sigma^2)$ so
$E(\varepsilon^3\mid T,X)=0$, and thus $E[\phi_a s_b]=0$.
Therefore $\phi_a$ is orthogonal to both the mean and variance nuisance tangent directions and is the efficient influence function for $\theta_a$ in this model.

Therefore, the semiparametric efficiency bound at $a$ is
\begin{equation}
\label{eq:varbound_global}
\mathcal V(a)
=\Var(\phi_a)
=\sigma^2\,E\!\left[\Big(\tfrac{k(T,X)}{f_{T,X}(T,X)}\Big)^2\right]
=\sigma^2\int\!\!\int \frac{k(t,x)^2}{f_{T,X}(t,x)}\,dt\,dx.
\end{equation}

Next, we we minimize $\mathcal V(a)$ over $a$ satisfying (R1) and (R2), equivalently over $k$ satisfying
\eqref{eq:k_constraints_global}. Introduce Lagrange multipliers $\lambda_0(x)$ for the pointwise constraint $\int k(t,x)dt=0$
and a scalar multiplier $\Lambda$ for the global constraint $\int\!\!\int t k(t,x)\,dt\,dx=1$.
The Lagrangian is
\[
\mathcal L
=\int\!\!\int \frac{k(t,x)^2}{f(t,x)}\,dt\,dx
+\int \lambda_0(x)\Big(\int k(t,x)\,dt\Big)dx
+\Lambda\int\!\!\int t\,k(t,x)\,dt\,dx -\Lambda.
\]
For each $(t,x)$, the first-order condition with respect to $k(t,x)$ is
\[
\frac{2k(t,x)}{f(t,x)}+\lambda_0(x)+\Lambda t=0
\quad\Rightarrow\quad
k(t,x)=-\tfrac12\big(\lambda_0(x)+\Lambda t\big)f(t,x) .
\]

Imposing the pointwise constraint $\int k(t,x)\,dt=0$ for each $x$,
\[
0
=\int k(t,x)\,dt
=-\tfrac12\lambda_0(x)\int f(t,x)\,dt
-\tfrac12\Lambda\int t\,f(t,x)\,dt.
\]
Let $\mu(x) = E[T|x]$ and note that
\[
\int f(t,x)\,dt = f_X(x),
\qquad
\int t\,f(t,x)\,dt = f_X(x)\mu(x),
\]
so
\[
-\tfrac12\lambda_0(x)f_X(x)-\tfrac12\Lambda f_X(x)\mu(x)=0
\quad\Rightarrow\quad
\lambda_0(x)=-\Lambda\mu(x)
\]
whenever $f_X(x)>0$ on the support.

Substituting back yields
\[
k(t,x)
=-\tfrac12\big(-\Lambda\mu(x)+\Lambda t\big)f(t,x)
=c\,(t-\mu(x))\,f_{T,X}(t,x),
\qquad c:=-\tfrac12\Lambda.
\]

The remaining constraint is
\[
1
=\int\!\!\int t\,k(t,x)\,dt\,dx
=c\int\!\!\int t(t-\mu(x))\,f_{T,X}(t,x)\,dt\,dx
=c\,E\big[T(T-\mu(X))\big].
\]
But
\[
E\big[T(T-\mu(X))\big]
=E\big[T^2\big]-E\big[T\mu(X)\big]
=E\big[\Var(T\mid X)+\mu(X)^2\big]-E\big[\mu(X)^2\big]
=E_X[\Var(T\mid X)],
\]
so
\[
c=\frac{1}{E_X[\Var(T\mid X)]}.
\]
Thus the minimizing $k$ is
\[
k^\star(t,x)
=\frac{(t-\mu(x))\,f_{T,X}(t,x)}{E_X[\Var(T\mid X)]},
\]
and the corresponding weight $a^\star$ is obtained by integrating $k^\star$
from the upper boundary of the support back to $t$:
\[
a^\star(t,x)
=\int_t^\infty k^\star(u,x)\,du
=
\frac{\displaystyle \int_t^\infty (u-\mu(x))\,f_{T,X}(u,x)\,du}
     {\displaystyle E_X[\Var(T\mid X)]}.
\]
This $a^\star$ satisfies (R1) and (R2) and is the unique minimizer of the variance bound \eqref{eq:varbound_global}.

Plugging $k^\star$ into \eqref{eq:varbound_global} yields
\begin{align*}
\mathcal V_{\min}
&=\sigma^2\,E\!\left[\Big(\tfrac{k^\star(T,X)}{f_{T,X}(T,X)}\Big)^2\right]
=\sigma^2\,E\!\left[\Big(\tfrac{(T-\mu(X))\,f_{T,X}(T,X)}{E_X[\Var(T\mid X)]\,f_{T,X}(T,X)}\Big)^2\right]\\
&=\sigma^2\,\frac{E[(T-\mu(X))^2]}{(E_X[\Var(T\mid X)])^2}
=\sigma^2\,\frac{E_X[\Var(T\mid X)]}{(E_X[\Var(T\mid X)])^2}
=\frac{\sigma^2}{E_X[\Var(T\mid X)]}.
\end{align*}

which concludes the proof. 
\end{proof}

\end{document}